\documentclass[12pt,preprint]{aastex}

\slugcomment{Submitted, \today}

\shorttitle{Sky Quality Near Eureka}
\shortauthors{Steinbring et al.}

\input epsf
\def\plotone#1{\centering \leavevmode
\epsfxsize=1.0\columnwidth \epsfbox{#1}}
\def\plotonenarrow#1{\centering \leavevmode
\epsfxsize=0.5\columnwidth \epsfbox{#1}}

\DeclareMathAlphabet{\mathscr}{OT1}{pzc}{m}{it}

\begin{document}

\title{Astronomical Sky Quality Near Eureka, \\in the Canadian High Arctic}

\author{Eric Steinbring\altaffilmark{1}, William Ward\altaffilmark{2} \& James R. Drummond\altaffilmark{3}}

\altaffiltext{1}{Herzberg Institute of Astrophysics, National Research
Council Canada, Victoria, BC V9E 2E7, Canada}

\altaffiltext{2}{Department of Physics, University of New Brunswick, Fredericton, NB E3B 5A3, Canada}

\altaffiltext{3}{Department of Physics and Atmospheric Science, Dalhousie University, Halifax, NS B3H 4R2, Canada}

\begin{abstract}

Nighttime visible-light sky brightness and transparency are reported for the Polar Environment Research Laboratory (PEARL), located on a 610-m high ridge near the Eureka research station, on Ellesmere Island, Canada. Photometry of Polaris obtained in $V$ band with the PEARL All Sky Imager (PASI) over two winters is supported by standard meteorological measurements and visual estimates of sky conditions from sea level. These data show that during the period of the study, October through March of 2008/09 and 2009/10, the sky near zenith had a mean surface brightness of $19.7~{\rm mag}~{\rm arcsec}^{-2}$ when the sun was more than $12\arcdeg$ below the horizon, reaching $20.7~{\rm mag}~{\rm arcsec}^{-2}$ during astronomical darkness with no moon. Skies were without thick cloud and potentially usable for astronomy 86\% of the time (extinction~$<2$ mag). Up to 68\% of the time was spectroscopic ($\leq 0.5$ mag), attenuated by ice crystals, or clear with stable atmospheric transparency. Those conditions can persist for over 100 hours at a time. Further analysis suggests the sky was entirely free of ice crystals (truly photometric) $48\pm3$\% of the time at PEARL in winter, and that a higher elevation location nearby may be better.

\end{abstract}


\section{Introduction}\label{introduction}

Winters in the High Arctic are cold, dry, and dark. In northern Canada, within the Territory of Nunavut, the coasts of the ice-locked eastern arctic archipelago combine these conditions with high terrain, providing locations potentially above much of the surface cloud and turbulence, and leading to the prospect of good observing sites for optical/near-infrared astronomy \citep[See][]{Steinbring2010}. One such site under study is the location of the Polar Environment Atmospheric Research Laboratory (PEARL) at $80\arcdeg$ north latitude, on Ellesmere Island. The PEARL facility is at an elevation 610 m on a ridge at the tip of the Fosheim Peninsula. It is accessible via a 15 km long road from the sea-level research base at Eureka, providing reliable logistical support: power, broadband satellite communications, an all-weather airstrip, and sea access in summer. Eureka is operated by the civilian weather service of the Canadian government, Environment Canada (EC), providing hourly meteorological data (air temperature, barometric pressure, relative humidity, wind speed and direction) and visual cloud-cover estimates.

With the onset of continuous polar night in early October, air temperature drops quickly at Eureka. It reaches an average of $-30$C within a few weeks, and by December is typically near $-40$C. A strong thermal inversion layer develops, with a peak $10$C warmer than sea level at a median height near 800 m, although it can be below 600 m elevation (lower quartile) and above 1200 m elevation (upper quartile), based on a 50-year climatology \citep{Lesins2009a}. It is already well known from visual sky-quality estimates obtained at Eureka that under these conditions skies are infrequently obscured by thick cloud. The mean precipitable water vapor column at sea level drops below 2 mm, freezing out into the form of ice crystals. Most often (just over 50\% of the time) a layer of this ``diamond dust" is suspended within the inversion layer, although it can precipitate from otherwise clear skies. As seen from PEARL, under these conditions Eureka appears blanketed in haze. Looking upwards, the opacity of crystals can be thin, producing visible halos for the moon and bright stars.

Observations with millimetre-wave radar combined with Laser Detection and Ranging (LIDAR) from sea level near Eureka provide the best available data for characterizing the size and vertical distributions of these boundary-layer ice crystals, differentiating them from mixed-phase water and ice clouds at higher altitudes by measuring their returned relative backscatter cross-sections and depolarization \citep{Bourdages2009}. These show ice crystals are typically 15-70 $\mu$m in cross section, and have vertical density distribution which decays exponentially with height (linearly with air pressure), dropping by two orders of magnitude from sea level to 1200 m \citep[Figure 7 in][]{Bourdages2009}. This is a reduction by a factor of 10 by 600 m, with a further factor of 2 decrease at 800 m elevation. That also correponds to roughly the elevations of the highest local terrain near Eureka, and events of wind-blown snow being the dominant component have been recorded \citep{Lesins2009b}. In fact, attaining the highest practical elevation, and rising above much of the ice crystal layer was an important aspect of siting PEARL, as its primary purpose is for the observation of chemical and dynamical signatures in the middle and upper atmosphere by optical means. An issue for astronomy though is the opacity of ice crystals and the fraction of time that the atmosphere above PEARL is subject to thin attenuation by any other contributors - one possibility being so-called ``arctic haze", aerosol pollutants transported from southern latitudes \citep[See][and references therein]{Quinn2007}. Quantification of transparency and sky brightness using an all-sky camera is desirable for comparison with other sites worldwide. Such an analysis for Mauna Kea using the CFHT SkyProbe was reported in \cite{Steinbring2009}.

The PEARL All Sky Imager (PASI) was deployed in 2007, and is suitable for an analysis similar to that of the Mauna Kea SkyProbe study. PASI \citep[See][for details of the instrument]{Veselinovic2008} was designed for the detection of atmospheric emission lines in airglow; primarily for the purpose of characterizing atmospheric waves (primarily tides and buoyancy waves - typically termed ``gravity waves" in the atmospheric community) and aurora, which can appear near the horizon, and sometimes overhead. It is an autonomous imager composed of an up-looking $1024\times1024$ CCD camera and a fisheye lens, yielding approximately square-$0.5\arcdeg$ pixels near zenith, viewing through a Plexiglass dome-shaped window in the ceiling of a warm room on the PEARL roof. Color information is provided by cycling through a filter wheel containing 5 narrowband (10~\AA-wide) filters. Although not selected for the purposes of astronomy, one of these has a central wavelength of 5725~\AA, comparable to $V$ band. This region was chosen to avoid airglow and aurora, providing a background sky measurement once every 15 minutes. This provides a useful cadence and a sufficiently deep exposure to image a few bright stars, including $\alpha$ Ursae Minoris (Polaris) which is just 10 degrees off zenith when viewed from Eureka.

This paper outlines our study of sky transparency and sky brightness at PEARL based on PASI measurements of extinction estimated from photometry of Polaris, correlated with meteorological and visual sky quality estimates from Eureka. The data and their reduction is outlined in Section~\ref{observations}, with further analysis of the fraction of time free of ice crystals follows in Section~\ref{analysis}, followed by a summary and conclusions in Section~\ref{summary}.

\section{Observations and Results}\label{observations}

The PASI data cover the periods from 1 October through 31 March during the winters of 2008/09 and 2009/10, which span 8733 hrs in total. Only PASI engineering data were acquired in 2007. Intermittent electrical faults beginning in 2010 lead to gaps in the data and none were obtained in the winter of 2010/11. Saturation of the PASI detector by bright moonlight restricted its operation during winter 2008/09 when 90 s exposures were used, and this was mitigated in winter 2009/10 by reducing the exposures to 45 s - although still not sufficiently to allow observation during all moon phases. The moon was above and below the horizon for almost equal parts, 4186 hrs and 4547 hrs, respectively. For brief periods the binning of the detector was changed, the focus of the lens altered, or the protective dome intentionally covered; all instances easily identified by inspection of stars in the frames, and those data excluded. Another issue is occasional ice buildup on the PASI dome, at times obscuring stars, and also leading to increasing scattered light due to the moon, and especially near sunrise. During the periods of the study the sun was above the horizon for 926 hrs and below for 7807 hrs. Nautical twilight begins on 3 October, and sun angles below $-12\arcdeg$ occur until mid-March, for a total of 5250 hrs. Astronomical darkness (sun below $-18\arcdeg$) occured during 3191 hrs of that time.

Although not intended to assess astronomical sky quality, hourly visual estimates of cloud cover are obtained from the weather station at Eureka which conform to a familiar set of conditions: perfectly clear, and more or less than 50\% clear sky. As discussed in \cite{Steinbring2010}, with the onset of darkness the predominant condition reported by the observer is `Ice crystals' (51\% of the time) which would include the reporting of any crystals seen suspended in the air. This can at times occur during conditions which would otherwise be considered `Clear' (15\%). The other bins are `Mostly clear' (6\%), `Mostly cloudy' (5\%), or `Cloudy' (4\%), with all others considered `Indeterminate' (18\%) including snow, blowing snow, rain, sleet or fog. No observations are obtained at midnight and 1 AM local time in Eureka. During the period of study there were 29062 PASI frames obtained. Each was correlated with the hourly observation obtained nearest in time from Eureka, for a total of 7266 hours of correlated observations (typically 4 PASI observations per hour). Having a visual check was helpful in characterizing the fraction of time that PASI observations were corrupted due to icing and scattered light. That was likely the case when a frame had no electronic errors, poor focus or vignetting, and skies were considered to be clear or subject only to ice crystals in Eureka.

\begin{figure}
\plotonenarrow{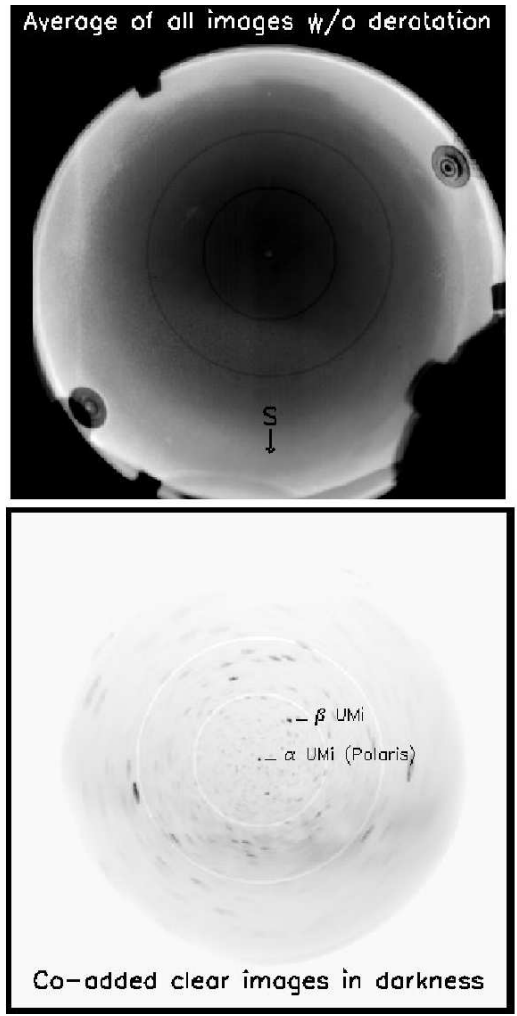}
\caption{Images obtained with the PASI system. Above are all images co-added; below are only those obtained in nautical twilight or darker, when the moon was below the horizon and the visual estimate from Eureka was of clear skies.}
\label{figure_images}
\end{figure}

Figure~\ref{figure_images} shows the co-addition of all PASI images. Some vignetting of the field due to other instruments deployed nearby PASI are evident, as is the effect of higher sky brightness associated with the sun and moon, both providing an effectively uniform illumination of the horizon in the average image. Circular fields spanning $80\arcdeg$ and $40\arcdeg$ centred on the north celestial pole are indicated by black circles. In the top panel, excess brightness within those rings is due to reflected light from ice accumulated on the dome itself, some of which could be scattered artificial light from Eureka, which lies to the southeast. Below is an image of the co-added frames when the sun was more than $12\arcdeg$ below the horizon, the moon was also below the horizon, and when visual inspection indicated `Clear' skies. Frames have been co-rotated. Sky brightness is noticeably reduced near zenith. Note the field distortion further than $40\arcdeg$ from the pole; with the 45 s to 90 s exposures also contributing to star trails. This is improved for those stars within $20\arcdeg$; the result for $\beta$ UMi (${\rm Declination}=+74\arcdeg9\arcmin$) is an image stretched over a few pixels. This effect is minimized for Polaris, which is just 44\arcmin~ from the north celestial pole, and therefore always falls within the same group of eight pixels on the detector. Polaris is a Cepheid variable, with ${<V>}=2.008$ mag \citep{Brown1994}, and is of interest due to its varying pulsation rate \citep{Turner2005}. Studies of it have used $\beta$ UMi ($V=2.07$ in the Hipparchos Catalog) as a non-variable comparison star, and we employ it here for obtaining relative photometry.

\begin{figure}
\plotone{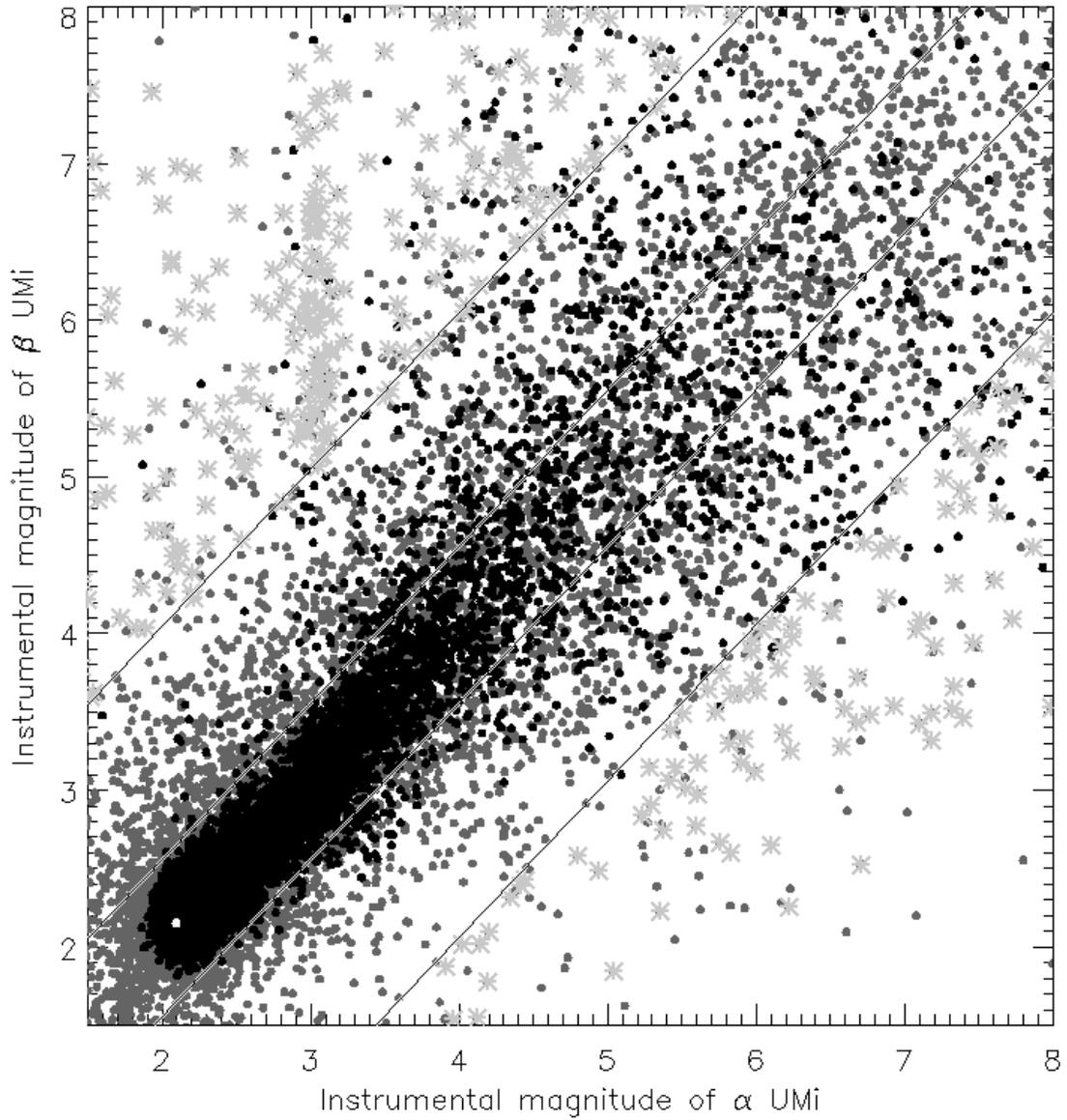}
\caption{Differential photometry of Polaris using $\beta$~UMi as a comparison star. Times when photometry differed by more than 2 magnitudes are indicated by gray crosses, likely caused by icing of the lens. Dark circles correspond to dark skies; a white dot indicates the intersection of the modes for both datasets.}
\label{figure_photometry}
\end{figure}

Synthetic aperture photometry ($1\arcdeg$ apertures, effectively $2\times2$ pixels), was carried out on Polaris and $\beta$~UMi using the IRAF task APER and custom scripts developed in Interactive Data Language. Sky subtraction was performed using $3\arcdeg$-wide annuli for each star, and a relative airmass correction was applied based on the hourly barometric pressure for dry air. No color correction has been made to transform to the standard Johnson filter set. The filter bandpass avoids strong telluric lines, and for Rayleigh scattering only, at $0.57\mu{\rm m}$ the mean atmospheric pressure of PEARL corresponds to an extinction of 0.099 mag at zenith using the prescription in \cite{Hayes1975}. Instrumental magnitudes are shown in Figure~\ref{figure_photometry}, which are fixed to an (outside the atmosphere) mean brightness for Polaris of 2.008 mag (zeropoint of $12.618$ mag) during dark, visually clear skies - with no additional attenuation by ozone or aerosols. It will be shown later that those remaining constituents can be quantified by inspecting the variation in stellar brightness, provided that the relative measurement has a stable baseline. The observational uncertainty per data point is dominated by sky background, which for purely Poisson statistics is 0.15 mag r.m.s. This is improved somewhat for those observed in darkness (0.13 mag), which are indicated by filled circles. The light grey crosses are those data which were affected by ice, obvious when the difference between Polaris and $\beta$ UMi exceeds 2 magnitudes, and especially in one case which leads to a vertical stripe of grey crosses at a nearly fixed brightness for Polaris. The intersection of the median photometry of both stars is indicated by the white dot.

\begin{figure}
\plotone{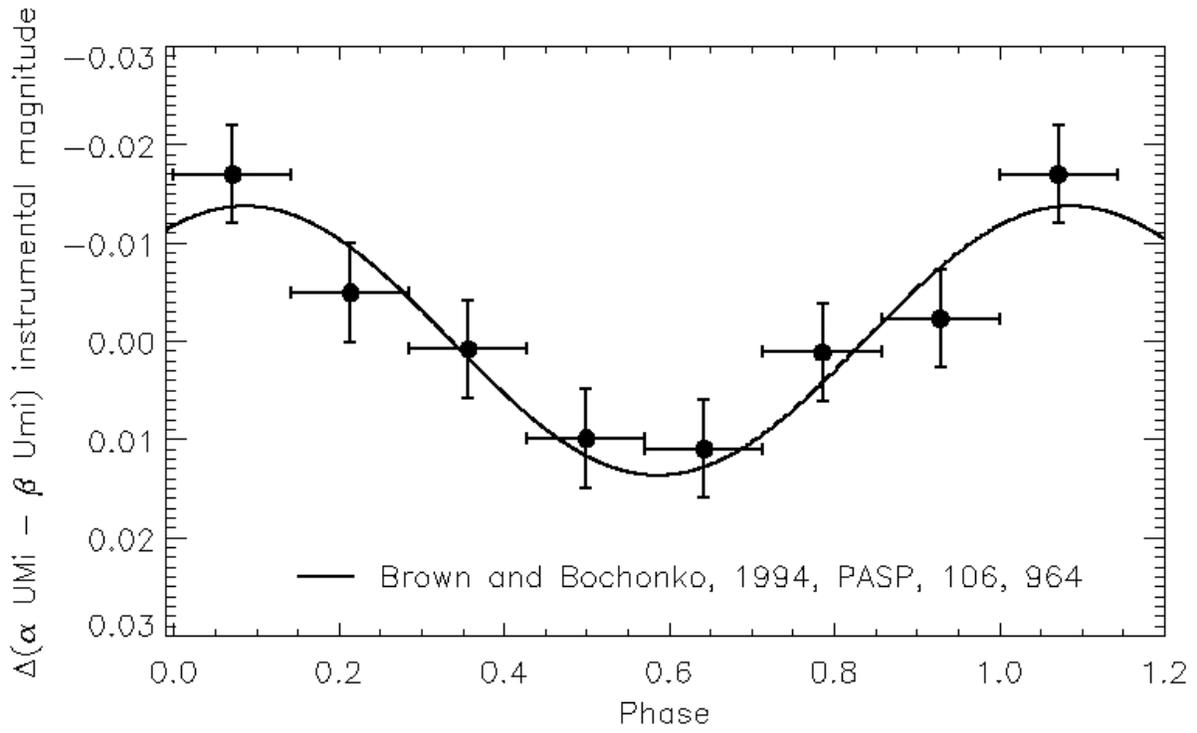}
\caption{Lightcurve of Polaris phased with a period of 3.97 days and divided into 7 equal phase bins. An offset of $\Delta (\alpha~{\rm UMi} - \beta~{\rm UMi})=0.062$ mag has been subtracted.}
\label{figure_lightcurve}
\end{figure}

After phase-wrapping according to the 3.96959-day period Polaris reported in \cite{Brown1994}, as per their ephemeris
$${\rm JD}_{\rm max light} = 2431495.91 + 3.96959E + 2.0116\times10^{-7}E^2,$$
the relative photometry of those observations obtained in darkness and visually clear skies, $\Delta (\alpha~{\rm UMi} - \beta~{\rm UMi})$ is shown again in Figure~\ref{figure_lightcurve} as instrumental magnitudes, with a constant offset of $0.062$ mag subtracted. The data have been binned to 7 equal phase bins, which for the 5148 observations suggests that the expected photometric uncertainty per bin is roughly $0.13/\sqrt{5148/7}=0.005$ mag. A sinusoidal model with a 3.96959-day period and depth of 0.014 mag from \cite{Brown1994} is overplotted. That this is resolved illustrates how, although the absolute PASI photometry of individual frames is noisy, relative photometry is stable enough to give confidence in the zeropoint at the 0.01-mag level over the course of the study.

\begin{figure}
\plotone{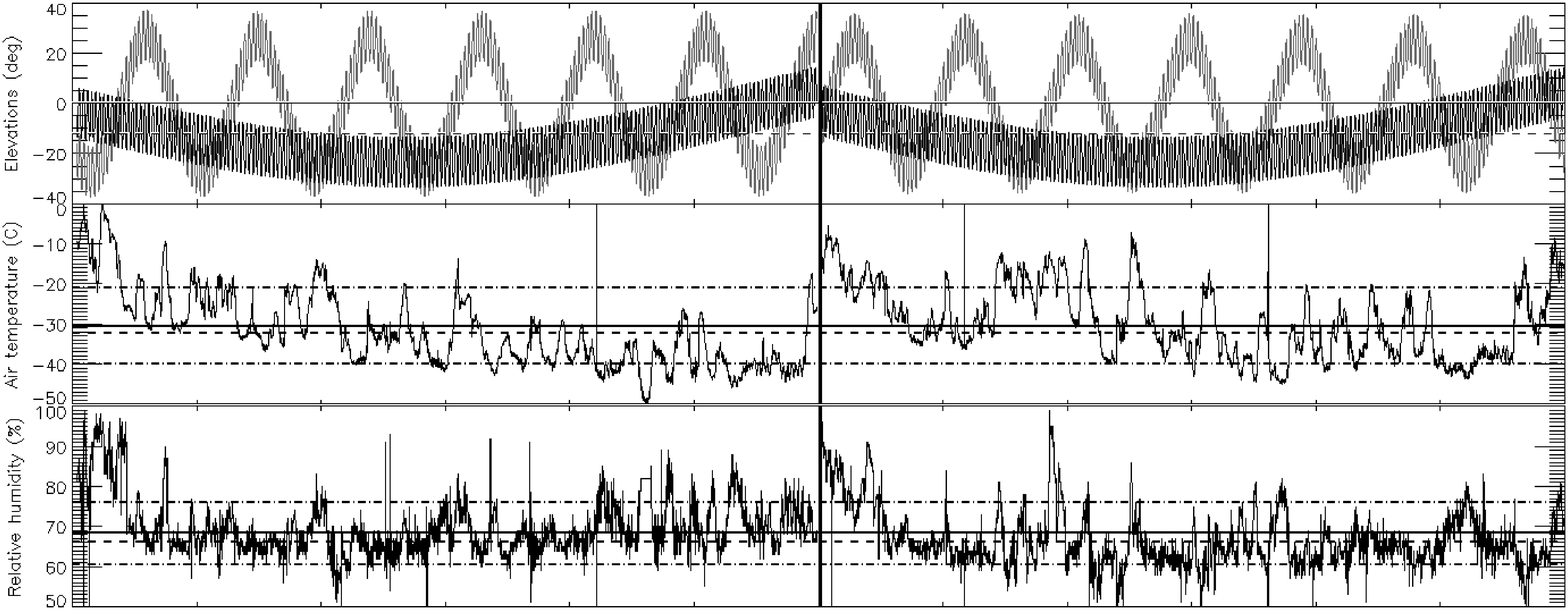}\\
\plotone{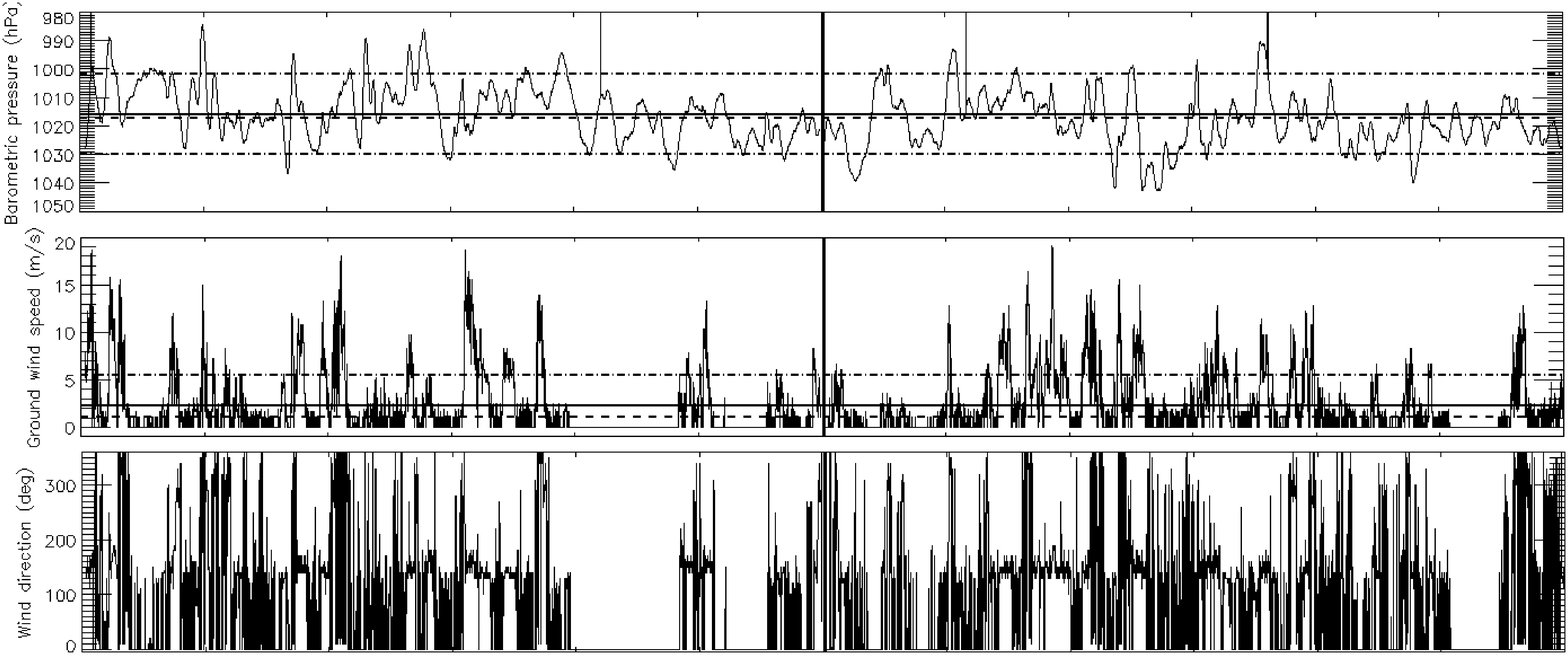}\\
\plotone{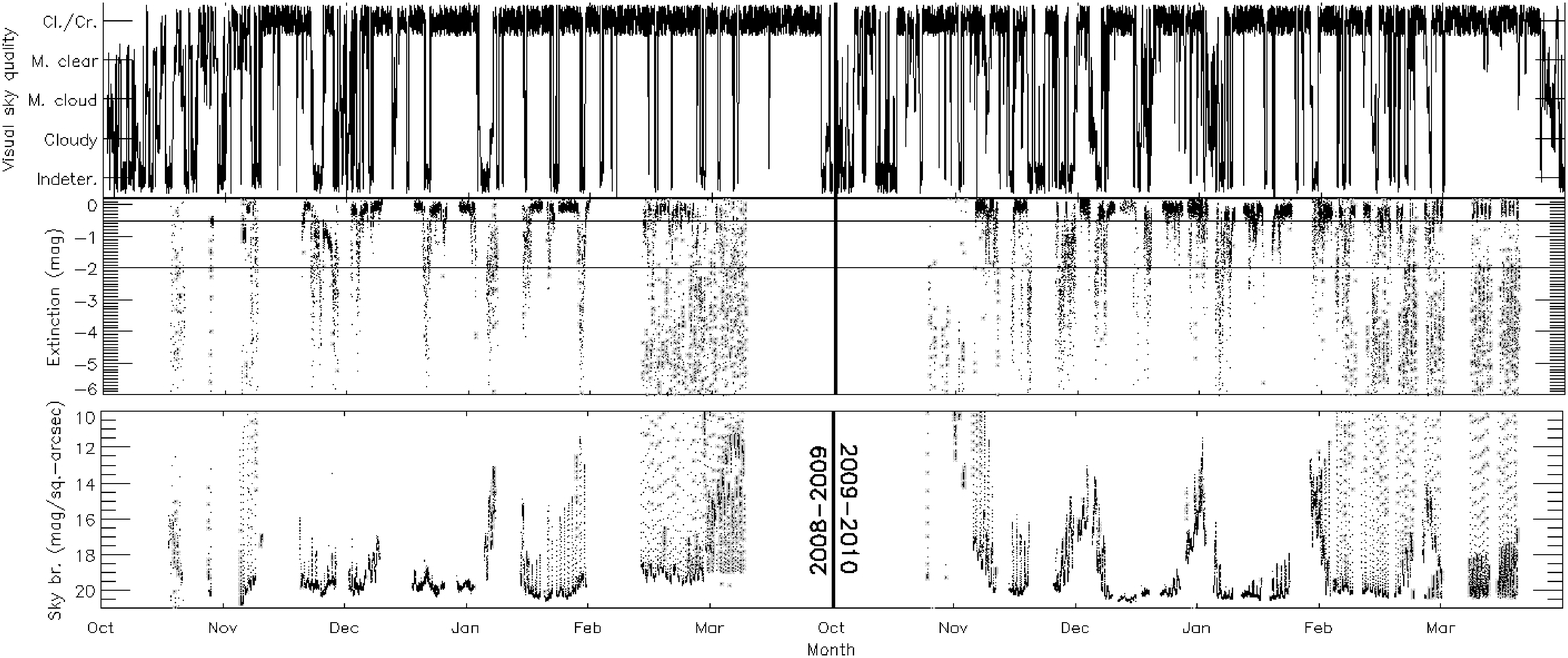}
\caption{Available data during the period of the study. At top are the elevations of the sun and moon. Below are all hourly average meteorological data for Eureka, at sea level: air temperature, relative humidity, barometric pressure, windspeed and direction, and visual sky-quality estimate. At bottom are the derived extinction and sky brightness from PASI.}
\label{figure_data}
\end{figure}

Subtracting this model Polaris lightcurve (for all data) reduces the result to an estimate of extinction, if there are no other sources of {\it added} light, for example enhanced apparent brightness due to an uneven background component. We will consistently refer to the result as extinction, presuming that this is due to attenuation by cloud, air, and any other atmospheric absorber. Extinction is shown at the bottom of Figure~\ref{figure_data}, which has a mode of 0.118 mag. At the top of the figure are the elevations of the sun and moon, and below that are the available meteorological data at sea level: air temperature, relative humidity, pressure, wind speed and direction, and the associated sky quality estimate. The consistently cold winter air leads to stable, saturated relative humidity. On the other hand, ozone concentration above Eureka is known to fluctuate due to atmospheric chemistry associated with the polar vortex, and is depleted following sunrise. Total vertical column densities were greater than 300 Dobson units ($8\times10^{18}~{\rm molecules}~{\rm cm}^{-2}$) in satellite and in-situ measurements taken during spring 2008 \citep{Batchelor2010}. In comparison with models for mid-latitude sites, this suggest a lower bound to extinction due to ozone at $0.57~\mu{\rm m}$ between $0.01$ mag \citep{Hayes1975} and $0.03$ mag \citep{Patat2011}. This is close to the limits of the photometric accuracy, but suggests that after accounting for Rayleigh scattering, only a small constituent of the remainder ($0.118-0.099=0.019$ mag) could be due to aerosols. An alternative would be that the atmospheric attenuation is underestimated (due to a systematic error in the photometry) but it will be seen later in Section~\ref{photometric} that the distribution in atmospheric attenuation makes that unlikely to be by more than about 0.03 mag. In short, there is no evidence in these data for strong attenuation by aerosols.

\begin{figure}
\plotone{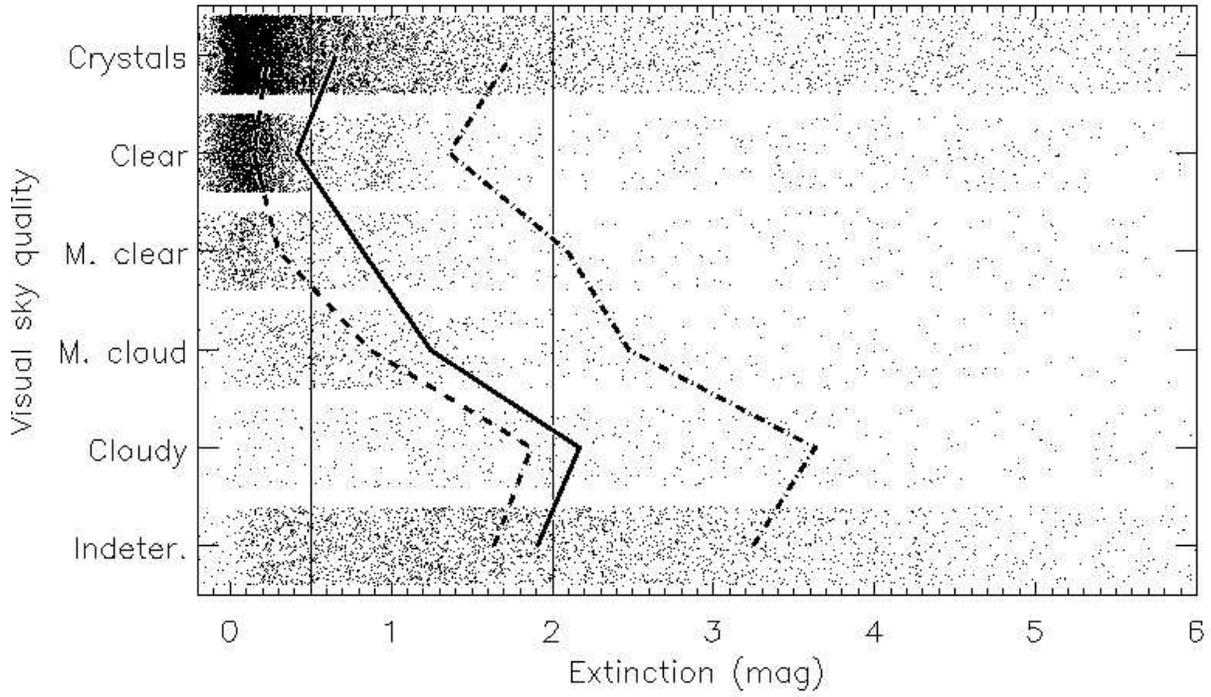}
\caption{Correlation of visual sky-quality estimates from Eureka and PASI data. Mean extinctions in each visual bin are connected by solid lines; dashed lines for median, dot-dashed for mean plus one standard deviation.}
\label{figure_correlation}
\end{figure}

A striking feature in the extinction measurements are periods lasting many days to weeks of low attenuation punctuated by a brief occurrences of high opacity lasting typically less than a few days. Figure~\ref{figure_correlation} helps illustrate the agreement between that and visual estimates of sky quality. Mean extinction for each of the visual bins of `Crystals', `Clear', `Mostly clear', `Mostly cloudy', `Cloudy', and `Indeterminate' are connected by solid lines; medians by dashed lines, and dot-dashed for mean plus one standard deviation from the mean. Random offsets within the visual-estimate bins have been applied to help show their relative distributions. 

Two useful divisions emerge in comparing visual estimates, PASI results, and expectations for astronomy. The first is a criteria for clear skies. The mean extinction is $0.71\pm0.24$ mag (1 $\sigma$); most conditions considered visibly clear or with crystals have extinction less than 0.5 magnitudes. It will be apparent later in the analysis that photometric accuracy of PASI makes the latter a natural limit to discrimination. Also, this amount of extinction can reasonably be considered to allow spectroscopic observations, and so we consider those of 0.5 mag or better as `Clear or spectroscopic.' Any choice involves some arbitrariness, of course, although less so as conditions degrade. For example, the Gemini observing condition constraints for Mauna Kea and Cerro Pachon use 1 mag $V$-band extinction as `cloudy' (revised from 2 mag prior to semester 2011A; see
http://www.gemini.edu/sciops/telescopes-and-sites/observing-condition-constraints), and above 3 mag as `unusable', when background becomes too detrimental at thermal infrared wavelengths, and reliable guiding is difficult. Skies considered `Mostly cloudy' or `Cloudy' in Eureka have PASI extinctions of 2 mag or more. Furthermore, by that definition, `Indeterminate' conditions are often cloudy - both the mean and median fall near 2 mag. So this too provides a useful limit, and conservatively falls within conditions considered astronomically useful elsewhere; those worse are deemed `Unusable.'

\begin{figure}
\plotone{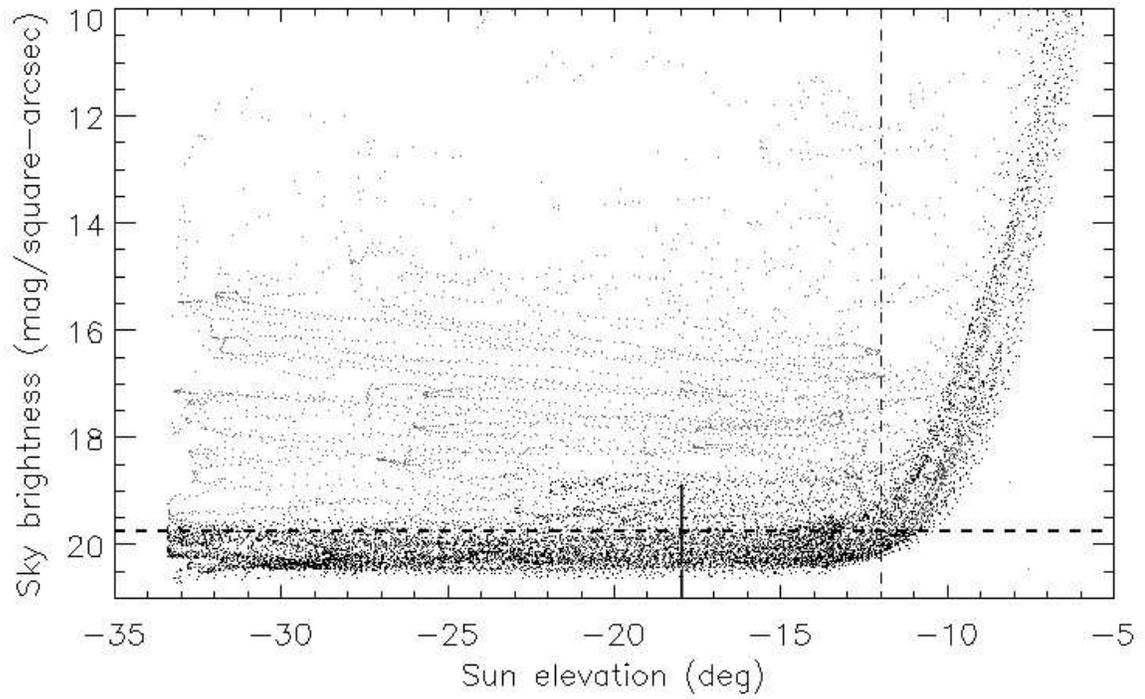}
\caption{Sky surface brightness in instrumental magnitudes as a function of sun elevation angle; mean observed sky brightness indicated by the horizontal dashed line. Sky darkness levels out for sun angles greater $-12\arcdeg$ (vertical dashed line), becoming flat before reaching $-18\arcdeg$ (short vertical line).}
\label{figure_sky_brightness}
\end{figure}

Another product of the PASI observations is an estimate of the sky surface brightness, although we caution that this is an instrumental magnitude for a non-standard filter. Figure~\ref{figure_sky_brightness} shows these plotted as a function of sun angle, falling to a mean of $19.7~{\rm mag}~{\rm arcsec}^{-2}$ when that is below $-12\arcdeg$, without any restriction on illumination by the moon. This sun angle is appropriate for estimating the fraction of time suitable for near-infrared observations. It corresponds approximately to `grey time' for Gemini (North and South) when skies reach $V>19.5~{\rm mag}~{\rm arcsec}^{-2}$. One would expect by analogy that in the visible this should also be comparable to conditions at Dome A, which is at $80\arcdeg$ south latitude. And indeed, it is similar to that of $19.8~{\rm mag}~{\rm arcsec}^{-2}$ reported by \cite{Zou2010} within the SDSS $i$-band filter of the Chinese Small Telescope Array (CSTAR) instrument. It can be seen in Figure~\ref{figure_sky_brightness} that the locus of points is not yet flat at $-12\arcdeg$, although that occurs before sun angles reach $-18\arcdeg$ (short vertical line). The darkest skies recorded with PASI were $20.7~{\rm mag}~{\rm arcsec}^{-2}$, which is consistent with independent measurements with a Unihedron sky-brightness monitor obtained in February 2011 ($21~{\rm mag}~{\rm arcsec}^{-2}$; Suresh Sivanandam, private communication). This is also the same as the median for Gemini, and comparable to the median of $20.5~{\rm mag}~{\rm arcsec}^{-2}$ reported for CSTAR under moonless clear nights. 

Possible contributions to sky brightness beyond those from the sun and moon are worth mentioning, although not discussed in detail here. The PASI instrument has filters designed to detect auroral emission, and that study will be reported elsewhere. As expected, there is no evidence of aurora in our data, either from visual inspection or from the photometry. The emission would need to fall within the bandpass, and be a bright sustained event near zenith to appear in Figure~\ref{figure_sky_brightness} as a vertical ``spike" for a fixed sun elevation. A small contributor may be artificial lights. The PEARL facility often has some external lamps on for safety reasons, and technicians are at times working on the PEARL roof with flashlights. The ``street" lights at the base in Eureka (15 km to the southeast) at night are also noticeable from PEARL. Observations with standard filters and away from PEARL could set more stringent limits on this. Also of interest would be estimates of sky brightness in the near and thermal infrared, which should be low, due to the cold ambient temperature. For now, the PASI sky-brightness measurements are helpful to understand limitations in the extinction data. Brighter skies will provide poorer photometry, with reduced contrast of Polaris against the background. Improved sensitivity comes with darker skies, which for the latitude of Eureka becomes optimal with sun elevations lower than $12\arcdeg$ below the horizon, without moonlight.

Table~\ref{table_samples} shows all PASI results and visual estimates taken nearest in time to them. The PASI results have been separated into periods when either the sun angle was higher than $12\arcdeg$ below the horizon, or the moon was above the horizon (independent of phase), called ``bright sky" and the opposite case, called ``dark sky." The fraction of time that skies are clear (extinction $\leq0.5$ mag) is between 61\% (dark sky) and 68\% (bright sky). This division provides a means of characterizing possible systematic errors in the clear-sky estimate, providing some limits on the range. At the high end, under bright sky the tendency would be to provide an overestimate of clear or spectroscopic conditions, if scattered light from crystals artificially enhanced the brightness of Polaris (produced a slight halo) over the uniform background, and so reduced its apparent extinction. This is not the case, as a large sky aperture was used, and it was found that shrinking the photometric aperture (to a single pixel) or enlarging it did not appreciably improve the $S/N$ of photometry. An opposing effect may come into play at the low end of clear-sky fraction. As the study periods started in October and ended in March, and are not directly symmetric with solar elevation, the windier (which we will show later, corresponds to cloudier) part of early winter is somewhat over-represented by the dark period, possibly causing it to be an underestimate. We do not consider a baseline of two years to be sufficient to disentangle the coincidence of new moon and high winds. In either case, the fraction of time estimated as usable for astronomy (extinction $\leq2$ magnitudes) is the same, 86\% based on this analysis. Indeterminate conditions, for example those frames corrupted by mechanical vignetting or with ice on the lens, are expected to be uniformly distributed with sky clarity, that is, uncorrelated with cloudiness. 

\section{Analysis and Discussion}\label{analysis}

The expectation for scarcity of thick cloud over Eureka in winter is borne out by the PASI results. Most of the time, skies are either clear or have thin attenuation by ice crystals. Following closely equation 1 from \cite{Steinbring2009}:
$$A=A_{\rm air} + A_{\rm cloud} = A_{\rm air, 1}Z + A_{\rm cloud, 1}T$$
where cloud thickness $T$ is analogous to airmass $Z$ and subscripts of 1 indicate
normalizations to $Z=1$ and $T=1$. Here, a contribution to air could be aerosols. It is the strong power-law nature of the occurrences of ice crystals that makes it possible to differentiate those from air, and estimate the fraction of time unaffected by attenuation due to cloud. A reasonable assumption is that in the absence of cloud the variance in $Z$ is Gaussian. Variation in $T$ plausibly follows a power law, with thicker clouds occurring less often. And from \cite{Steinbring2009} it is expected that 
$$T=1-\alpha \log(\delta/\bar{\delta})$$
where $\alpha$ is a constant, and $\delta$ is the duration of cloud, with $\bar\delta$ indicating the mean. Qualitatively, this suggests that occurrences of thick clouds should be rare, while thinner clouds are exponentially more common.

\begin{figure}
\plotone{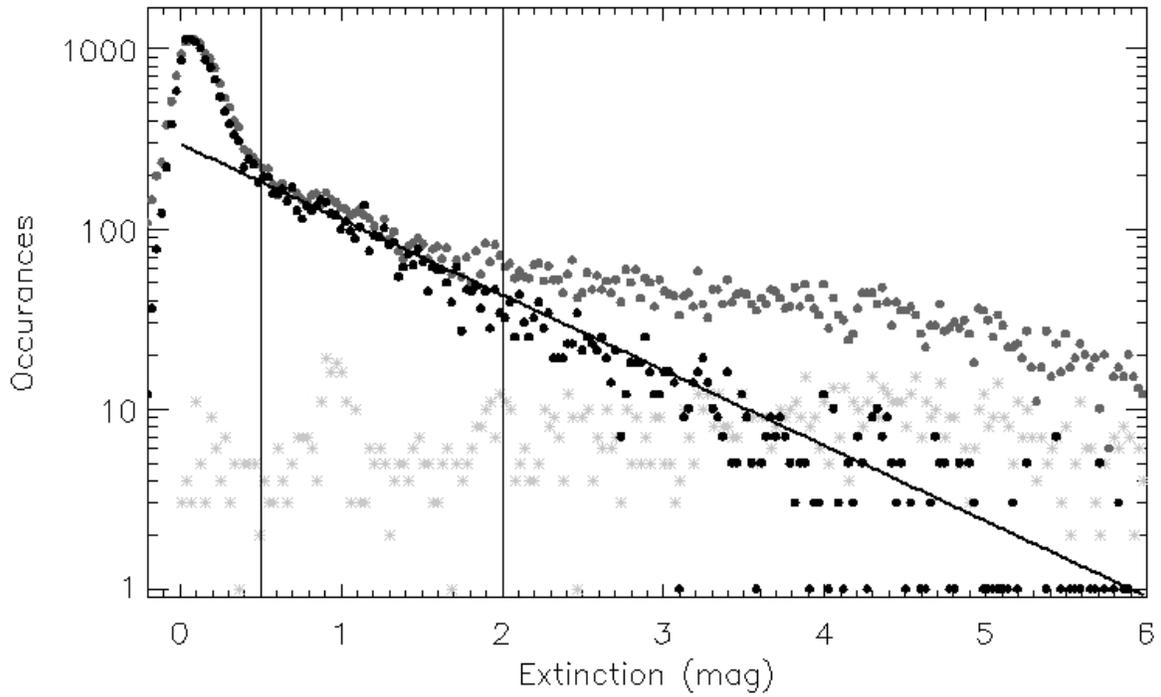}
\caption{Histogram of all extinction data. A power-law fit to between 0.5 and 2 magnitudes is shown extended to lower and higher values by a thick solid line. Those data obtained in darkness are indicated by the dark circles; frames when ice is suspected to have covered the lens are indicated by the gray crosses.}
\label{figure_histogram}
\end{figure}

A histogram of all extinction data is shown in Figure~\ref{figure_histogram}. Filled circles represent data taken during darkness. A power-law fit is shown, selecting only data with extinction greater than 0.5 mag and less than 2 mag. The slope is $-1.04\pm0.02$. Following the analysis in \cite{Steinbring2009}, $\alpha=-1/({\rm slope}\times A_{\rm cloud,1})=-1/(-1.04\times0.71)=1.35$. Note how well this curve fits, well beyond 3 magnitudes of extinction during dark skies when photometry is best. Frames which are believed to be affected by ice on the lens are indicted by the gray crosses. It is evident that these are uniformly distributed with extinction, and are most likely the cause of enhanced counts of extinctions greater than 2 magnitudes.

\subsection{Photometric Fraction of Time}\label{photometric}

\begin{figure}
\plotone{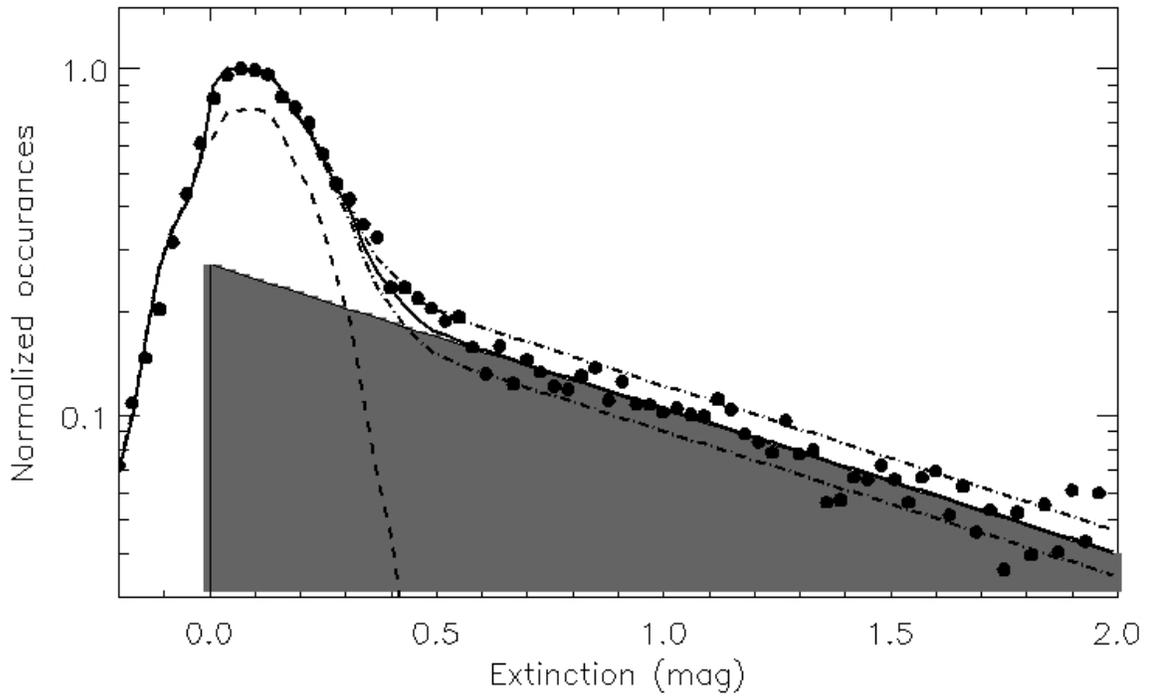}
\caption{Same as Figure~\ref{figure_histogram} except restricted to extinctions less than 2 magnitude. Attenuation by cloud (shaded region) combined with that from air (dashed curve) results in a model of the data (thick solid curve); dot-dashed curves indicate increasing or decreasing the estimate of clear skies by 3\%.}
\label{figure_model}
\end{figure}

Figure~\ref{figure_histogram} is a duplicate of Figure~\ref{figure_model}, with the difference that it is restricted to extinctions less than 2 mag and in bins of 0.03 mag. The model is shown in Figure~\ref{figure_model}, the combination of a power-law slope and a Gaussian distribution of $\sigma=0.13$ mag centred at 0.10 magnitudes. After the subtraction of ice crystals, the remainder should be due entirely to the observational scatter about the atmospheric attenuation. No other component seems necessary to allow a good fit. It may even be that a small ``shoulder" in the distribution of extinctions less than zero is reproduced with this two-component model. Although counterintuitive, negative measures of extinction are expected due to photometric scatter, because that inflates the width of the distribution in atmospheric attenuation (dashed curve). Increasing the atmospheric attenuation by 0.06 mag (shifting all data to the right in this diagram, but not changing the slope of the power law) is inconsistent with the combined attenuation of air-plus-cloud being greater than that for clouds only, which is nonphysical. However, simply fluctuating the ratio of air to cloud-model by 3\% yields the dot-dashed curves; which are consistent with both the slope of the power law and the mode of the atmospheric attenuation - providing some confidence in that measure, as discussed in Section~\ref{observations}. The integral under the dashed curve yields the truly photometric fraction of time, that is without any cloud or ice crystals, $48\pm3$\% as observed with PASI. This is similar to the 51\% fraction of time with $i$-band attenuation less than 0.11 mag reported for Dome A using CSTAR \citep{Zou2010}. It is somewhat poorer than the 56\% photometric fraction estimated for Mauna Kea in \cite{Steinbring2009} but still comparable, and based on the same method.

\subsection{Duration of Dark, Clear-Sky Periods}\label{duration}

\begin{figure}
\plotone{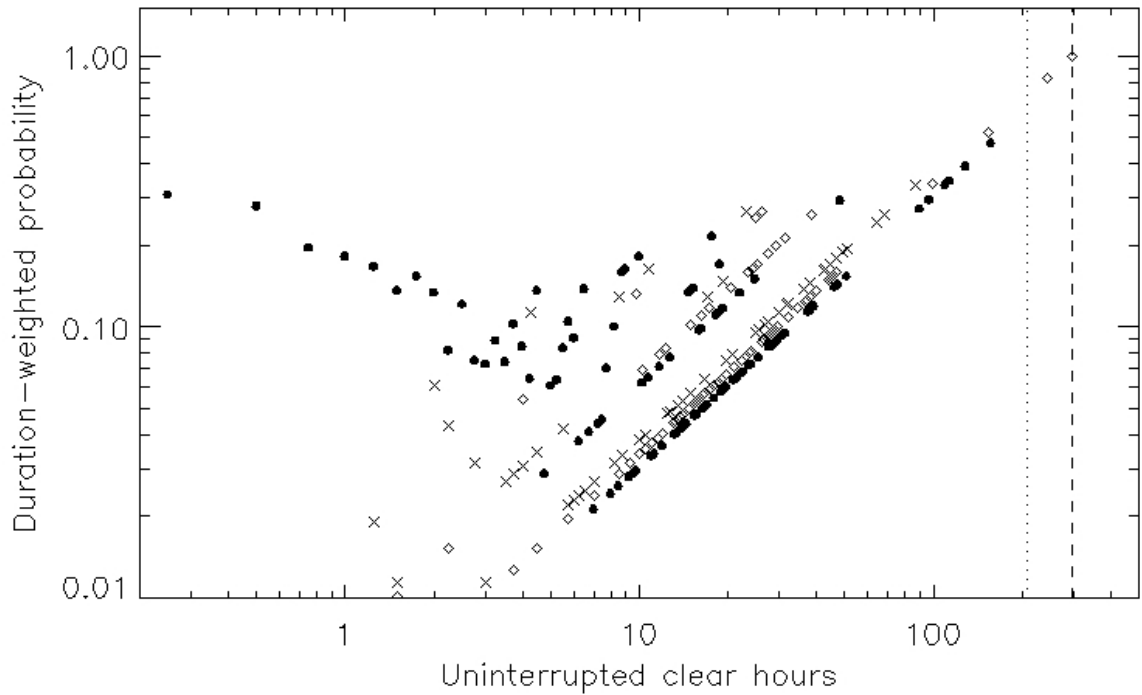}
\caption{Uninterrupted periods of clear skies in darkness, ideal cases (diamonds) are restricted to 300 hours and no moon (dashed line). Observed uninterrupted periods for PASI (dark circles) are limited by the longest period of uncorrupted frames (vertical dotted line). At least four periods over 100 hrs long were observed, for a total of 490 hrs; nearly three weeks.}
\label{figure_clear_skies}
\end{figure}

Long, uninterrupted periods of dark, clear skies are common at PEARL, and as expected, are as long or longer than those observed at sea level. Figure~\ref{figure_clear_skies} shows the duration-weighted probability of uninterrupted periods when skies which were both dark and had conditions of `Clear' and `Ice crystals' had ice crystals as viewed in Eureka (crosses) or had extinction $\leq 0.5$ mag with PASI (filled circles). In those cases where an observation was not obtained in Eureka, the missing observation is assumed to be the condition immediately prior to the interruption. For comparison, the ideal theoretical uninterrupted PASI durations are shown as diamonds, which reaches a limit of 300 hrs for the moon conditions discussed above. The plotted PASI durations are conservative, including disruptions due to corrupted frames or ice on the lens window. The longest possible duration without being affected by those errors is approximately 200 hrs. At least 4 cases of uninterrupted observations longer than 100 hrs were observed, surpassing the best for Eureka. The expected longest duration of clear observations would be $\delta_0=f_{\rm clear}*t_{\rm obs}=0.45*200=90$ hrs. The PASI data indicate that the average duration of periods when extinction was less than 0.5 mag was between 50 and 60 hours, which is an estimate of $\bar\delta$, the mean duration for cloud of thickness $T=1$.

\subsection{Wind, Local Terrain and Ice Crystals}\label{dependence_on_wind}

Although strong winds occur at Eureka, winds are usually light, and so the average windspeed is low. The mean windspeed during the study period was 2.2 ${\rm m}~{\rm s}^{-1}$. One caveat is that the Eureka instrument does not report windspeeds in excess of 20 ${\rm m}~{\rm s}^{-1}$, which may skew the mean downwards. Another is that the instrument is not sensitive to winds below 1 ${\rm m}~{\rm s}^{-1}$, recording these as 0 ${\rm m}~{\rm s}^{-1}$. The fraction of time that winds are observed to be calm ($<1~{\rm m}~{\rm s}^{-1}$) is 42\%, and interestingly, this is similar to the photometric fraction of time at PEARL based on PASI. It is important to emphasize that the latter is a statistical estimate based on the ensemble of attenuation measurements; the accuracy of individual measurements is not sufficiently good to discriminate at the 0.1-mag level for instances of 0 ${\rm m}~{\rm s}^{-1}$ windspeed.

\begin{figure}
\plotone{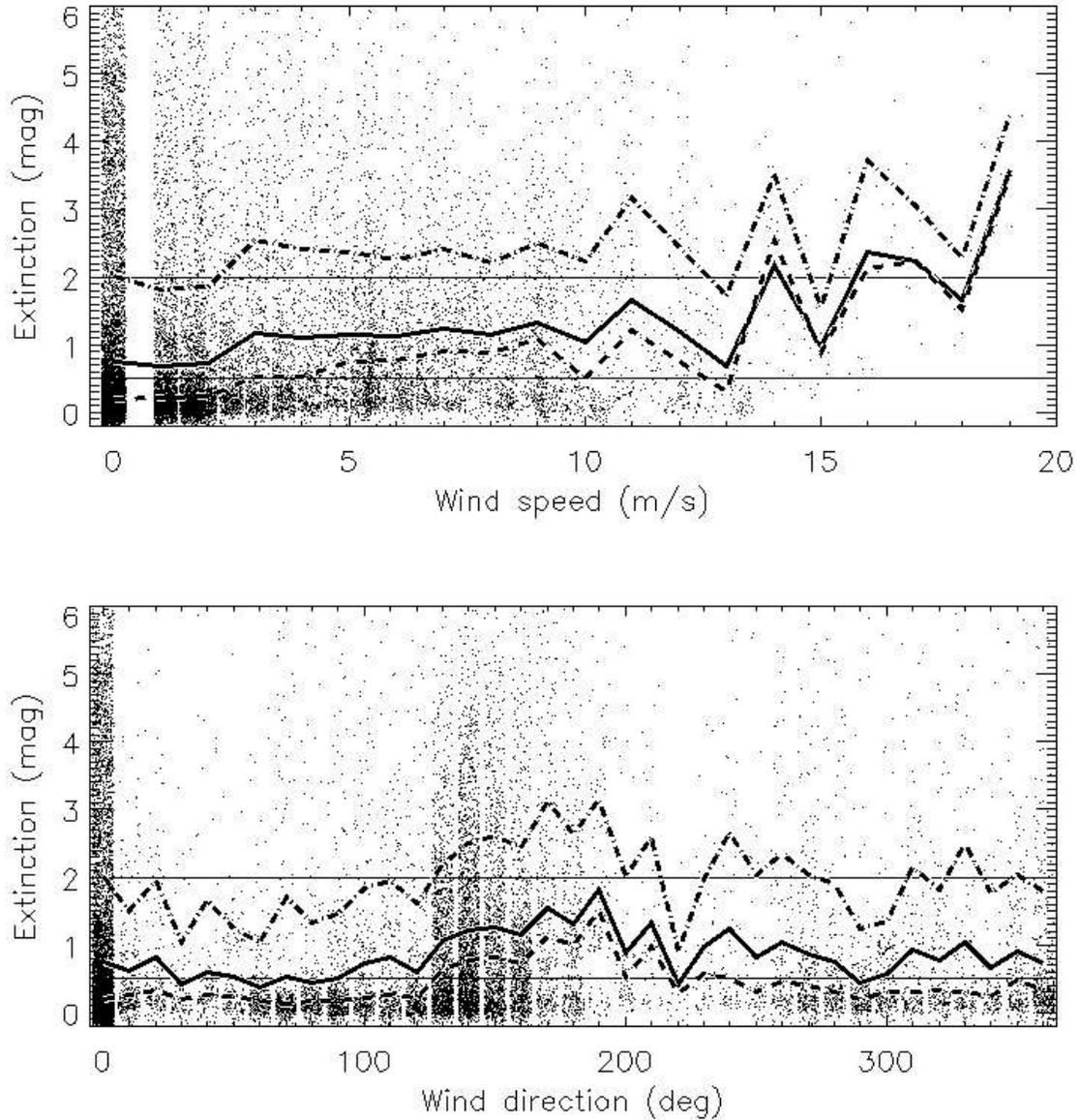}
\caption{Distribution of extinction with wind speed (top) and direction (bottom). The Eureka anemometer is not sensitive to winds less than 1 ${\rm m}~{\rm s}^{-1}$, and records them as 0 ${\rm m}~{\rm s}^{-1}$. The azimuthal direction is given as East of North, with $0\arcdeg$ indicating windspeed of 0 ${\rm m}~{\rm s}^{-1}$, and $360\arcdeg$ indicating North. Solid lines indicate the means; dashed, medians; and dot-dashed, mean plus one standard deviation. There is a trend towards cloudy conditions when winds are high in Eureka and from the southeast.}
\label{figure_wind}
\end{figure}

There is, however, strong evidence that clear skies at PEARL are associated with calm winds. Figure~\ref{figure_wind} shows the distributions of extinction with wind speed (top) and wind direction (bottom) as recorded in Eureka. The mean and standard deviation of extinction are maximal for wind speeds over 12 ${\rm m}~{\rm s}^{-1}$ or winds from the southeast to west. These may be the consequence of high winds blowing ice crystals from higher terrain, as discussed in \cite{Lesins2009b}. 
There are a few elevated prominences (by 10 m or 20 m) less than 0.5 km along the PEARL ridge, mostly to the west, and approximately 30 km to the east is Blacktop Ridge, which reaches 825 m. The minimum extinction (a median of 0.1 mag) occurs for wind speeds below 1 ${\rm m}~{\rm s}^{-1}$, as one would expect for extinction typically dominated by attenuation by air. It may also correspond to the condition of true in-situ diamond dust, without influence of blown ice crystals.

\subsection{Optical Depth and Barometric Pressure}\label{relation_to_barometric_pressure}

\begin{figure}
\plotone{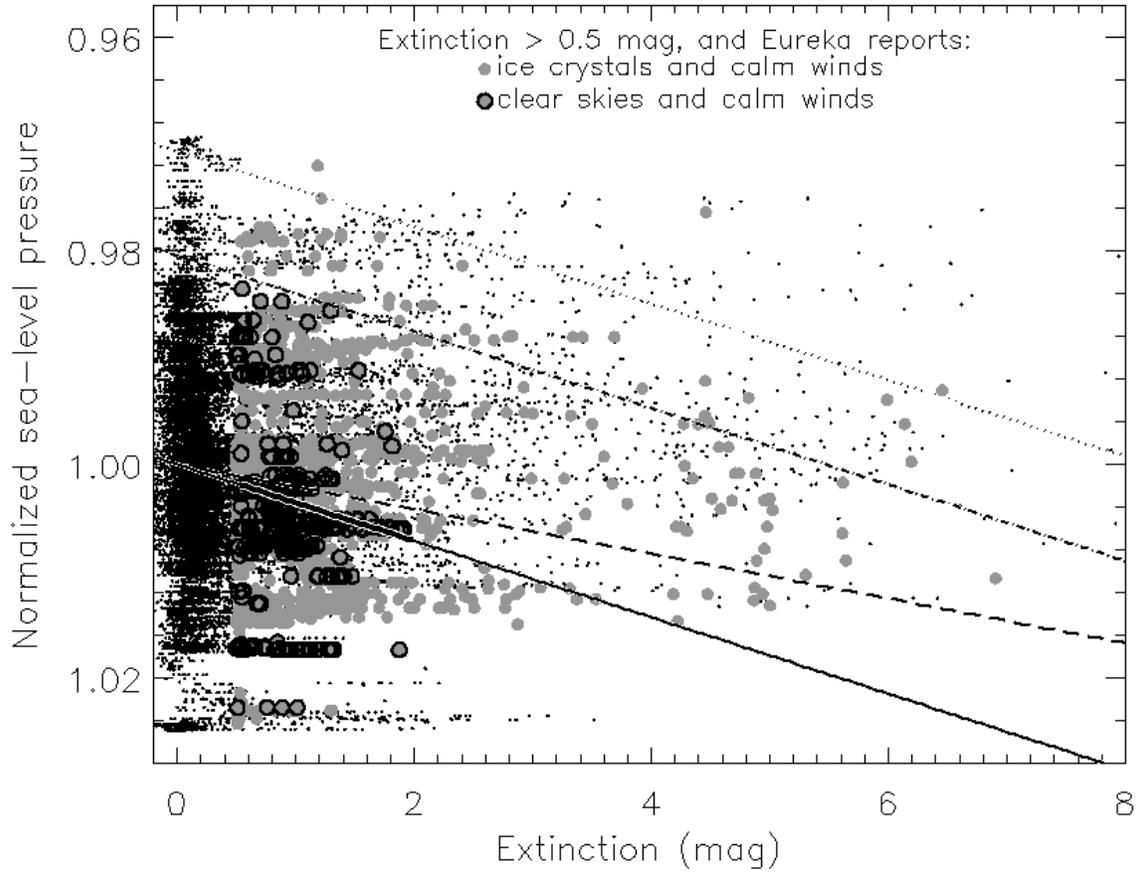}
\caption{Plot of extinction as a function of barometric pressure, relative to the mean at sea level. Highlighted are cases of extinction $>0.5$ mag when Eureka reports that winds are calm and is either subject to ice crystals or has clear skies. A fit to those last data (thick solid line) is shown, extended to lower and higher extinctions, with parallel lines drawn at the 2-$\sigma$ pressure (dot-dashed) and 3-$\sigma$ pressure (dotted). A model relating cloud thickness $T$ to pressure $p$, is plotted as a thick dashed line. See text for details.}
\label{figure_crystals}
\end{figure}

A vertical density distribution of ice crystals linear with air pressure is anticipated (see Section~\ref{introduction}), and so a correlation between extinction of ice crystals and barometric pressure was looked for. Figure~\ref{figure_crystals} shows extinction measured with PASI, plotted relative to normalized sea-level barometric pressure. Considering only extinctions $>0.5$ mag eliminates influence from atmospheric attenuation. A further restriction to calm conditions avoids confusion with wind-blown crystals or stormy weather, for which low pressure may presage cloudy skies. The resulting subsample is broken down into two categories: when Eureka reported either ice crystals or clear skies. The latter should presumably select cases of thin attenuation by ice crystals in PASI data. A linear fit to those data is shown as a thick solid line, which has a slope of $3.01\times10^{-3}~{\rm mag}^{-1}$, or $3.06~{\rm mag}~{\rm hPa}^{-1}$ at 1017 hPa.  This fit is extended to lower and higher extinctions by a thin solid line. It may be that the remaining data obtained under windless conditions also reflect this trend, and to help illustrate that, parallel lines are shown, offset by 2 $\sigma$ (dot-dashed) and 3 $\sigma$ (dotted) from the mean pressure. It is interesting that the distribution of ice crystals falls off above (lower pressures than) the 2-$\sigma$ line, even though the data are not scarce there; only a few instances of ice crystals occur beyond the 3-$\sigma$ line.

Increasing ice crystal attenuation with pressure might be explained by a linear relationship with $T$, that is,
$$T\propto p/\bar{p}.$$ And following from $T=1-\alpha\log{(\delta/\bar{\delta})}$ the relative change in thickness with pressure would then be
$$\Delta T = (p_{2.0} - p_{0.5})/{\bar{p}}=\alpha\log{(\delta_{0.5}/\delta_{2.0})},$$
where the subscripts 0.5 and 2.0 indicate those attenuations in magnitudes. Choosing these two points is only for the convenience of obtaining the ratio of $\delta_{0.5}/\delta_{2.0}$ as those relative fractions of time can readily be obtained from Table~\ref{table_samples}, giving a constant of proportionality of $1.35\log{(68/14)}=2.13$. A thick dashed line using this slope is overplotted in Figure~\ref{figure_crystals}. This proportionality is not as steep as the fitted slope, but has the same sense: an increase in pressure prescribes an upper limit to the thickening of cloud. Put another way, for a given pressure a thicker cloud must have a shorter duration, which for regular samples corresponds to fewer detections. And so the distribution of ice crystal events one expects to find is sparser towards high extinction, and is bounded by a upper limit proportional to pressure. This is consistent with the observations of ice crystals seen in Figure~\ref{figure_crystals}.

A plausible scenario is that when winds are calm in Eureka, which from Section~\ref{dependence_on_wind} is taken to be 42\% of the time, the thickness of the layer of ice crystals simply increases with atmospheric density due to a uniform distribution of ice crystals in air. This correlation is poorer when there is wind. As mentioned above, it may be that ice crystals under those conditions are also blown from terrain, complicating the picture. When calm, it is consistent with PEARL being at sufficiently high elevation to at times rise above the layer of ice crystals. From Figure~\ref{figure_crystals}, a drop of 3\% in barometric pressure relative to mean station pressure (939 hPa) would correspond to consistently clear skies (always $A<0.5$ mag) at PEARL. Interestingly, a pressure drop of 3\% (to 912 hPa) also corresponds to an increase of effective altitude of approximately 225 m for dry air, which relative to the 610 m elevation of PEARL is a height of 835 m above sea level. Furthermore, if at that pressure height the photometric fraction of time were also increased in the same proportion as clear skies, that is, growing at minimum by $100/68=1.47$, then this would occur at least $1.47\times42\%=62\%$ of the time.

\section{Summary and Conclusions}\label{summary}

We have presented an analysis of stellar photometry using an all-sky camera at PEARL and supported by meteorological and visual sky-quality estimates from Eureka, over two consecutive winters beginning in 2008. The observed PASI lightcurve of Polaris is of sufficient quality to discern its 28 millimag, 4 day periodicity, and provides two basic outcomes. First, it confirms skies are dark in the visible, $19.7~{\rm mag}~{\rm arcsec}^{-2}$ on average for all sun elevations below $-12\arcdeg$, and reaching $20.7~{\rm mag}~{\rm arcsec}^{-2}$ for sun elevations below $-18\arcdeg$ when there is no moon. Further observations in standard filters would be helpful to characterize sky brightness in the infrared, and due to the contribution of scattered artificial light from the facility itself or the nearby Eureka base. Second, PEARL is at a high enough elevation to be above the bulk of the ice crystal boundary layer in winter. For 86\% of the time this has an extinction less than 2 mag, suitable for astronomical observations. Up to 68\% of the time extinctions are less than 0.5 mag. This spectroscopic/clear-sky fraction is superior to higher elevation sites in southern Canada, for example, those reported (by different methods) for Mount Kobau \citep[60.5\%;][]{Brosterhus1972} and similar mountains in the southern British Columbia interior \citep{Hickson1989}. That these conditions can persist for over 100 hours at a time allows for observations that would be impossible from a single mid-latitude site (not part of a global network).

Further analysis of the data show that the truly photometric fraction of time (without any cloud) is $48\pm3$\%, which may be comparable to some of the best sites worldwide; Dome A in Antarctica, and Mauna Kea in Hawaii. The attenuation by air only, in instrumental magnitudes, was $0.12\pm0.03$ mag, which suggests only a small contribution from arctic haze is possible. The uncertainties are limited by photometry with PASI. Best conditions at PEARL are associated with calm winds. A possible trend in extinction with barometric pressure suggests that an increase in elevation may appreciably reduce optical depth of ice crystals. According to this analysis, an elevation increase of 225 m would effectively ensure skies are clear when winds are calm in Eureka. One location nearby Eureka in principle providing such an elevation advantage is the top of Blacktop Ridge, and so this may provide encouragement to further explore high terrain near Eureka for candidate sites.

\acknowledgements

The assistance of Tony Xie in preparing the PASI data is much appreciated, as is the Canadian Network for the Detection of Climate Change and their technicians for support in deployment and operation of the instrument at PEARL. We thank Ray Carlberg and Greg Fahlman for helpful conversations. An anonymous referee provided thoughtful comments which lead to improvements on the original manuscript. Eureka meteorological data were provided by Environment Canada through the National Climate Data and Information Archive.

\begin{deluxetable}{lrr}
\tablecaption{Contemporaneous Visual and All-Sky Camera Observations\label{table_samples}}
\tablewidth{0pt}
\tabletypesize{\small}
\tablehead{\colhead{} &\colhead{} &\colhead{Fraction}\\
\colhead{} &\colhead{} &\colhead{of total}\\
\colhead{Description of Sample} &\colhead{Quantity} &\colhead{(\%)}}
\startdata
Visual Estimates                               &      &   \\
Crystals                                       &3873  &53\phantom{\tablenotemark{b}}\\
Clear                                          &1275  &18\phantom{\tablenotemark{b}}\\
Mostly clear                                   &429   &6\phantom{\tablenotemark{b}}\\
Mostly cloudy                                  &283   &4\phantom{\tablenotemark{b}}\\
Cloudy                                         &185   &2\phantom{\tablenotemark{b}}\\
Indeterminate\tablenotemark{a}                 &1217  &17\phantom{\tablenotemark{b}}\\
\cline{1-3}
Total                                          &7266  &100\phantom{\tablenotemark{b}}\\
\\
All-Sky Camera Frames                          &      &    \\
Bright sky: & & \\
~~~~~~~~Clear or spectroscopic ($\leq0.5$~mag) &5872  &68\tablenotemark{b}\\
~~~~~~~~Intermediate ($>0.5$~mag, $\leq2$~mag) &1562  &18\tablenotemark{b}\\
~~~~~~~~Unusable ($>2$~mag)            &1235  &14\tablenotemark{b}\\
\cline{2-3}
~~~~~~~~Subtotal                               &8683  &30\phantom{\tablenotemark{b}}\\
\\
Dark sky: & & \\
~~~~~~~~Clear or spectroscopic ($\leq0.5$~mag) &5874  &61\tablenotemark{b}\\
~~~~~~~~Intermediate ($>0.5$~mag, $\leq2$~mag) &2452  &25\tablenotemark{b}\\
~~~~~~~~Unusable ($>2$~mag)            &1323  &14\tablenotemark{b}\\
\cline{2-3}
~~~~~~~~Subtotal                               &9671  &33\phantom{\tablenotemark{b}}\\
\\
No measurement\tablenotemark{c}                &3458  &12\phantom{\tablenotemark{b}}\\
Corrupted\tablenotemark{d}                     &4419  &15\phantom{\tablenotemark{b}}\\
Ice on lens\tablenotemark{e}                   &2831  &10\phantom{\tablenotemark{b}}\\
\cline{1-3}
Total                                          &29062 &100\phantom{\tablenotemark{b}}\\
\enddata
\tablenotetext{a}{Includes unobserved and those during snow, blowing snow, rain, sleet or fog.}
\tablenotetext{b}{Quoted as percent of subtotal.}
\tablenotetext{c}{Includes those with mechanical vignetting and electronic errors, or possibly incorrect focus; photometry is indeterminant, skies may be cloudy.}
\tablenotetext{d}{As with 'No measurement', except skies were reported to be visually clear or with ice crystals in Eureka.}
\tablenotetext{e}{From inspection of image or photometry; either one of the two bright stars not visible, or change in extinction from one frame to the next exceeds 2 magnitudes during a period when skies were visually clear or with ice crystals.}
\end{deluxetable}

\begin{thebibliography}{}

\bibitem[Batchelor et al.(2010)]{Batchelor2010} Batchelor, R.L., Kolonjari, F., Lindenmaier, R., Mittermeier, R.L., Daffer, W., Fast, H., Manney, G., Strong, K. \& Walker, K. 2010, Atmos. Meas. Tech., 3, 51

\bibitem[Bourdages et al.(2009)]{Bourdages2009} Bourdages, L., Duck, T.J., Lesins, G., Drummond, J.R. \& Eloranta, E.W. 2009, Atmos. Chem. Phys., 9, 6881

\bibitem[Brosterhus et al.(1972)]{Brosterhus1972} Brosterhus, E., Pfannenschmidt, E. \& Younger, F. 1972, \jrasc, 66 (1), 1

\bibitem[Brown \& Bochonko(1994)]{Brown1994} Brown, C.F. \& Bochonko, R. 1994, \pasp, 106, 964

\bibitem[Hayes \& Latham(1975)]{Hayes1975} Hayes, D.S. \& Latham, D.W. 1975, \apj, 197, 593

\bibitem[Hickson \& Bennett(1989)]{Hickson1989} Hickson, P. \& Bennett, P.D. 1989, \jrasc, 83 (2), 122

\bibitem[Lesins et al.(2009a)]{Lesins2009a} Lesins, G., Duck, T.J. \& Drummond, J.R. 2009, Ocean-Atmosphere, incomplete reference

\bibitem[Lesins et al.(2009b)]{Lesins2009b} Lesins, G., Bourdages, L., Duck, T.J., Drummond, J.R., Eloranta, E.W. \& Walden, V.P. 2009, Atmos. Chem. Phys., 9, 1847

\bibitem[Patat et al.(2011)]{Patat2011} Patat, S. Moehler, K. O’Brien, E. Pompei, T. Bensby, G. Carraro, A. de Ugarte Postigo, A. Fox,
I. Gavignaud, G. James, H. Korhonen, C. Ledoux, S. Randall, H. Sana, J. Smoker, S. Stefl \& T. Szeifert, 2011, \aa, 527, 91

\bibitem[Quinn et al.(2007)]{Quinn2007} Quinn, P.K., Shaw, G., Andrews, E., Dutton, E.G., Ruoho-Airola, T., \& Gong, S.L., 2007, Tellus B, 59 (10), 99

\bibitem[Steinbring et al.(2009)]{Steinbring2009} Steinbring, E., Cuillandre, J.-C. \& Magnier, E. 2009, \pasp, 121, 29

\bibitem[Steinbring et al.(2010)]{Steinbring2010} Steinbring, E., Carlberg, R., Croll, B., Fahlman, G., Hickson, P., Leckie, B., Pfrommer, T. \& Schoeck, M. 2010, \pasp, 122, 1092

\bibitem[Tarasick \& Bottenhiem(2002)]{Tarasick2002} Tarasick, D.W. \& Bottenheim, J.W. 2002, Atmos. Chem. and Phys., 2, 197

\bibitem[Tomasi et al.(2010)]{Tomasi2010} Tomasi, C., Petkov, B., Stone, R.S., Benedetti, E., Vitale, V., Lupi, A., Mazzola, M., Lanconelli, C., Herber, A. \& von Hoyningen-Huene, W. 2010, Journal of Geophysical Research, 115, D02205

\bibitem[Turner et al.(2005)]{Turner2005} Turner, D.G., Savoy, J., Derrah, J., Abdel-Sabour Abdel-Latif, M. \& Berdnikov, L.N. 2005, \pasp, 117, 207 

\bibitem[Veselinovic(2008)]{Veselinovic2008} Veselinovic, D. 2008, ``Atmospheric Gravity Waves Detected in the Arctic Atmosphere by an All Sky Airglow Imager" MSc Thesis, University of New Brunswick

\bibitem[Zou et al.(2010)]{Zou2010} Zou, H., Xu, Z., Zhaoji, J., Ashley, M., et al. 2010, \aj, 140, 602 

\end{thebibliography}
\end{document}